# ML and AI for density functional theory: different priorities for Kohn-Sham and orbital-free DFT, for electronic and nuclear DFT

Xin-Hui Wu,[a,1] Sergei Manzhos [b,2]

[a] Department of Physics, Fuzhou University, Fuzhou 350108, Fujian, China
[b] School of Materials and Chemical Technology, Institute of Science Tokyo, Ookayama 2-12-1, Meguro-ku, Tokyo 152-8552, Japan

## Abstract

We overview similarities and, importantly, differences in computational bottlenecks and accuracy requirements that can be addressed with machine learning (ML) and artificial intelligence (AI) techniques in electronic and nuclear DFT. From these follow different promising methodological and algorithmic choices depending on whether one machine learns the exchange correlation (XC) functional, the kinetic energy functional (KEF), the density or the basis functions. In particular, while the popular deep neural networks remain a potent choice in the context of KS DFT, we highlight their disadvantages when building KEFs and highlight conceptual advantages – yet to be fully realized - of symbolic regression for both electronic and nuclear DFT. We point out promising approaches that can be carried from the more extensively investigated ML-enhanced electronic DFT to nuclear DFT.

## 1 Introduction

Density functional theory (DFT) has firmly established itself as a practical solution to quantum many-body problems in situations where the solution of many-particle equations describing the "ab initio world" is impractical. We call the "ab initio world" a set of elemental particles and interactions considered. We consider two such worlds: one contrived for the description of atoms, molecules, and chemical compounds under Born-Oppenheimer approximation, and one contrived for the description of self-bound nuclei and nuclear matter. In the former case, the world consists of pointwise particles with electrostatic charges and masses mimicking nuclear cores and electrons. In the absence of relativistic effects (one often used approximation at the basis of the field of ab initio chemistry), elementary interactions include Coulombic interactions *only*. In addition, the electrons are treated as fermions with spin. In the case of clamped nuclei (another often used approximation at the basis of the field of ab initio chemistry) this world is described by the Schrödinger equation for the electrons. Despite the simplicity of the interactions considered, the solution of the Schrödinger equation in principle allows computing energies, structures (via energy minimization with respect to nuclei positions), electronic / band structures, bonding mechanisms, electronic excitations, and other properties of molecules and compounds with, in principle, any accuracy, barring non-BO and relativistic effects. At the same time, despite the simplicity of the interactions involved, the high-dimensional nature (the number of degrees of freedom is $3N_e$, where $N_e$ is the number of electrons) of the resulting Schrödinger equation makes its solution impractical except for smallest atoms and molecules due to strongly non-linear or combinatorial (depending on the method) scaling of CPU cost with system size.

[1] Correspondence: E-mail: wuxinhui@fzu.edu.cn
[2] Correspondence: E-mail: sergei.manzhos@mct.isct.ac.jp

In the case of nuclear physics, the ab initio world consists of nucleons, treated as massive fermions with spin and isospin degrees of freedom. These particles, which consist of quarks and gluons, interact through the strong nuclear force, supplemented by the Coulomb interaction. Since low-energy quantum chromodynamics is nonperturbative, practical ab initio nuclear theory does not usually start directly from quarks and gluons, but from effective interactions among nucleons. The nuclear Hamiltonian therefore contains nucleonic kinetic energy, two-nucleon interactions, and, in general, three- and higher-body interactions among nucleons. While the form of the kinetic energy operator (KEO) is the same for electrons and nucleons, in contrast to the Coulomb interaction in the electronic problem, nuclear forces are short-ranged, strongly spin- and isospin-dependent, and contain central, tensor, spin-orbit, and many-body components. Another essential distinction is that nuclei are self-bound systems, with no external potential corresponding to the clamped nuclei in molecular electronic-structure theory. The corresponding Schrödinger equation, or relativistic many-body equation for nuclei, in principle contains the information needed to compute nuclear binding energies, radii, shell structure, spectra, transitions, reactions, fission barriers, the nuclear-matter equation of state, etc. In practice, its direct solution is very restricted, because of the rapidly increasing dimensionality of the many-nucleon Hilbert space, the complexity of nuclear interactions, and the importance of many-body correlations.

In both the electronic and the nuclear many-body problem, DFT offers a practical way to obviate the dimensionality scaling issue due to the number of particles by reducing the many-body problem formulated in terms of N-body wave functions to the one-body level with the density distributions, whereby the energy is considered as a functional of the density – electron density in the case of electronic many-body systems or nucleon density in the case of nuclear many-body systems. The availability of such density functional would allow linear scaling of the cost with system size. In the absence of accurate universal density functionals, the idea of Kohn-Sham auxiliary system of non-interacting particles for many years offered a practical way to do DFT calculations with accuracy that is limited yet *sufficient* for most applications. KS DFT uses effective single-particle equations with effective potentials, within a mean-field paradigm. Despite the one-particle nature of the KS equation, the dimensionality scaling of computational cost with system size is formally cubic due to the need to compute many solutions of an eigenproblem. The effective potential is more complex than the interactions in the original Schrödinger equation. Exchange and correlation (XC) effects are subsumed into XC functionals that become a key accuracy limitation. At the same time, XC functionals, if sufficiently accurate, allow approaching the construction of a density functional by focusing on the modeling of non-interacting kinetic energy of the KS system. This is a key concept in orbital-free (OF) DFT pursued for both the electronic and nuclear problems. The accuracy of the kinetic energy functional (KEF) then becomes a key bottleneck.

The use of machine learning (ML), and recently of methods such as automatic formula discovery that move the field closer to the realm of artificial intelligence (AI), has recently become popular in addressing both accuracy and CPU cost bottlenecks in both electronic and nuclear DFT, and in both KS and OF paradigms. The targets of such methods often involve XC and KE functionals (to alleviate accuracy limitations), density (electron or nucleon density, respectively, for the electronic and nuclear problem) – whose prediction have the value of improving both accuracy and speed of calculations, and basis sets for computational efficiency. While targets such as functionals or density are common for the electronic and nuclear problem, and while certain qualitative criteria of accuracy are similar such as the reproduction of the shell structure of atoms and nuclei with ML functionals, quantitative and some qualitative criteria are different and some applications are specific, e.g. ML pseudopotentials for electronic DFT or symmetry-restoration in nuclear DFT. A similar set of ML/AI methods have found use in all these applications, ranging from neural networks to kernel methods to trees to symbolic regression. However, the requirements for accuracy and for computational cost of ML/AI solutions are quite different depending on whether one considers electronic or nuclear DFT, KS or OF DFT.

In this Perspective, we analyze similarities and differences in computational bottlenecks and accuracy requirements that can be addressed with machine learning and artificial intelligence techniques used for electronic and nuclear DFT in both Kohn-Sham and orbital-free formulations. We overview key ML/AI targets and methods used in different flavors of DFT and show that from these follow different promising methodologic and algorithmic choices depending on whether one machine learns the exchange correlation (XC) functional, the kinetic energy functional (KEF), the density (electron density or nucleon density), basis functions, or pseudopotentials (effective interaction). We only briefly summarize the principles and key equations of electronic and nuclear DFT to the extent necessary to highlight their conceptual differences and differences that follow in how they can be improved with ML/AI or data-driven approaches in general. Excellent comprehensive reviewers are available of both KS[1,2] and orbital-free DFT[3] for the electronic problem as well as for the nuclear DFT.[4–8] We will also not introduce relevant machine learning and AI methods, as abundant literature is available.[9–14] We do not consider delta learning of the results (e.g. of energies, geometries) of more accurate methods from DFT results or methods that bypass DFT by learning its outcomes from descriptors of chemical composition and structure,[15–18] but consider instead how ML/AI methods can be used to improve the DFT engine itself. We focus on more or less promising ML/AI approaches for these different flavors of DFT. In particular, we suggest that while the popular deep neural nets remain a potent choice in the context of KS DFT, their disadvantages may become bottlenecks when building KEFs for OF-DFT. We highlight the promise of non-NN methods for both electronic and nuclear DFT, and promising approaches that can be carried from the relatively more mature ML-enhanced electronic DFT to nuclear DFT. We also highlight conceptual advantages – although yet to be fully realized - of symbolic regression for both electronic and nuclear DFT, especially in the OF paradigm.

## 2 Similarities and differences between electronic and nuclear DFT in Kohn-Sham and orbital-free paradigms

### 2.1 *Brief summary of electronic DFT*

DFT for the electronic problem computes the energy of a system of $N_a$ ions (positioned at $\boldsymbol{R_i}$ in the Cartesian space $\boldsymbol{r}$) and $N_e$ electrons as functional of electron density[19] $\rho(\boldsymbol{r} \in R^3)$,

$$E = E[\rho(\boldsymbol{r})] = F[\rho(\boldsymbol{r})] + \int \rho(\boldsymbol{r}) v_{ion}(\boldsymbol{r}) d\boldsymbol{r} \tag{2.1.1}$$

where in Eq. (2.1.1), $E$ is further expanded into the contribution due to the energy of Coulombic interactions between the electron density and the ionic cores (or pseudocores in the case of the use of pseudopotentials) and the contribution due to electron-electron interactions that can be related to the electron wavefunction $\Psi(\boldsymbol{r}_e \in R^{3N_e})$ (not computed in DFT): $F[\rho(\boldsymbol{r})] = \min_{\Psi \to \rho} \langle \Psi | \hat{T} + \hat{V}_{ee} | \Psi \rangle$, where $\hat{T} = \nabla^2_{\boldsymbol{r}_e}$ is the kinetic energy operator, and $\hat{V}_{ee}$ is the operator for Coulombic electron-electron repulsion.[20] In the Kohn-Sham paradigm,[21] the electron density is represented as a sum of one-electron densities that are in turn represented as square moduli of wavefunctions called orbitals (in the following, we use atomic units unless stated otherwise and neglect spin and partial occupancy without loss of generality):

$$\rho(\boldsymbol{r}) = \sum_{i=1}^{N_e} \rho_i(\boldsymbol{r}) = \sum_{i=1}^{N_e} |\psi_i(\boldsymbol{r})|^2$$
$$\int d\boldsymbol{r} \rho_i(\boldsymbol{r}) = 1 \tag{2.1.2}$$

where the integral is over the extent of the system. The orbitals are solutions of the Kohn-Sham equation[21]

$$\left(-\frac{1}{2}\nabla^2 + v_{eff}(\boldsymbol{r})\right)\psi_i(\boldsymbol{r}) \equiv \hat{H}_{KS}\psi_i(\boldsymbol{r}) = \epsilon_i \psi_i(\boldsymbol{r}) \tag{2.1.3}$$

where $\epsilon_i$ are orbital energies (giving rise to band structure) and $v_{eff}(\boldsymbol{r})$ is the Kohn-Sham effective potential that incorporates, besides Coulombic interactions, a term accounting for exchange and correlation – the exchange-correlation functional $V_{XC}[\rho(\boldsymbol{r})] = \frac{\delta E_{XC}[\rho]}{\delta \rho(\boldsymbol{r})}$ – that makes $\rho(\boldsymbol{r})$ of Eq. (2.1.2) equal to the true electron density of the physical system:

$$v_{eff}(\boldsymbol{r}) = \sum_{i=1}^{N_a} \frac{Z_i}{|\boldsymbol{r} - \boldsymbol{R}_i|} + \int \frac{\rho(\boldsymbol{r}')d\boldsymbol{r}'}{|\boldsymbol{r} - \boldsymbol{r}'|} + V_{XC}[\rho(\boldsymbol{r})] \equiv V_{ion} + V_{Harteee} + V_{XC} \tag{2.1.4}$$

As the exact form of $V_{XC}[\rho]$ (or $E_{XC}[\rho]$)) is not known, its approximation constitutes a major *accuracy* limitation (regardless of computational cost) and has been subject to much research that resulted in a large number of available XC functionals that offer different trade-offs and are attractive for different applications. The KS equation is solved by representing the orbitals in a basis $\{\theta_k(\boldsymbol{r})\}$ and converting the KS differential equation into a matrix eigenvalue problem:

$$\begin{gathered} \psi_i(\boldsymbol{r}) = \sum_{k=1}^{N_{basis}} c_{ik}\theta_k(\boldsymbol{r}) \\ \boldsymbol{Hc} = \epsilon \boldsymbol{Sc} \\ \boldsymbol{H_{ij}} = \langle \theta_j(\boldsymbol{x}) | \hat{H}_{KS} | \theta_i(\boldsymbol{x}) \rangle \\ \boldsymbol{S_{ij}} = \langle \theta_j(\boldsymbol{x}) | \theta_i(\boldsymbol{x}) \rangle \end{gathered} \tag{2.1.5}$$

This step is the major *computational* bottleneck of KS DFT that is responsible for the cubic scaling of computational cost with large CPU cost prefactors, which makes it impractical to do calculations beyond about $O(10^3)$ atoms unless specialized methods and additional approximations are used.[22–26] This bottleneck can be said to be computational since with certain type of basis functions (in particular, plane waves), uniform convergence of the basis expansion is achievable. A real space grid can also be used.[27] Other popular types of basis sets, in particular more CPU-efficient LCOA (linear combination of atomic orbitals)-like bases and contracted bases used, in particular, in large-scale calculations,[24,28] also pose a basis optimization problem and therefore an *accuracy* limitation that could be addressed with ML. These basis functions can be numeric rather than analytic.[22] As $v_{eff}(\boldsymbol{r})$ depends on the density, the KS equation is solved self-consistently, and the number of self-consistent field (SCF) cycles and their convergence can constitute a numeric bottleneck. Once the KS equation is solved, the energy is expressed as

$$
\begin{aligned}
E &= \frac{1}{2}\sum_{i=1}^{N_{el}} \int d\boldsymbol{r}\psi_i^*\nabla^2\psi_i + \int d\boldsymbol{r}V_{ion}(\boldsymbol{r})\rho(\boldsymbol{r}) + \frac{1}{2}\int \frac{\rho(\boldsymbol{r}')\rho(\boldsymbol{r})d\boldsymbol{r}d\boldsymbol{r}'}{|\boldsymbol{r}-\boldsymbol{r}'|} + E_{XC}[\rho] \\
&\equiv E_{kin} + E_{ion} + E_{Hartree} + E_{XC} \\
&= \sum_{i=1}^{N_e} \epsilon_i - E_{Hartree} + E_{XC} - \int \frac{\delta E_{XC}[\rho(\boldsymbol{r})]}{\delta\rho(\boldsymbol{r})}\rho(\boldsymbol{r})d\boldsymbol{r}
\end{aligned}
$$

(2.1.6)

In the case of periodic system, the KS equation is solved at different points in the Brillouin zone (i.e. points sampling it), and Eq. (2.1.6) is integrated over Brillouin zone (with sampling points serving as quadrature grid).

Eq. (2.1.1) expresses orbital-free DFT. Minimization of $E$ with respect to electron density under an approximation for $F[\rho(\boldsymbol{r})]$ can be done with linear scaling and with small CPU cost prefactors, opening a way to routine calculations with ~$10^6$ atoms. The form of $F[\rho(\boldsymbol{r})]$ is a key *accuracy* bottleneck, and we will argue below that it also risks becoming a *computational* bottleneck with some ML methods. The search for $F[\rho(\boldsymbol{r})]$ (also called universal functional) is greatly facilitated by the vast experience of KS DFT development and application. Specifically, recognizing that (i) only the kinetic energy $E_{kin} = \frac{1}{2}\sum_{i=1}^{N_{el}} \int d\boldsymbol{r}\psi_i^*\nabla^2\psi_i$ in Eq. (2.1.6) depends on the orbitals, (ii) existing XC functionals are sufficiently accurate (i.e. possess practically impactful descriptive and predictive power) for most classes of materials, most OF DFT research represents $F[\rho(\boldsymbol{r})]$ as

$$
F[\rho] = E_{kin}[\rho] + E_{Hartree}[\rho] + E_{xc}[\rho]
$$

(2.1.7)

and approximations are constructed for $E_{kin} = \int \tau(\boldsymbol{r})d\boldsymbol{r}$, where $\tau(\boldsymbol{r})$ is KS kinetic energy density (KED) $\tau_{KS}(\boldsymbol{r}) = -\frac{1}{2}\sum_{i=1}^{N} \psi_i^*(\boldsymbol{r})\Delta\psi_i(\boldsymbol{r})$ or (most often) its positive-definite version $\tau_+(\boldsymbol{r}) = \frac{1}{2}\sum_{i=1}^{N_{el}} |\nabla\psi_i(\boldsymbol{r})|^2 = \tau_{KS}(\boldsymbol{r}) + \frac{1}{4}\Delta\rho(\boldsymbol{r})$. Approximations for $E_{kin}[\rho]$ may also include neglected correlations in KS kinetic energy. $E_{kin}[\rho]$ is as of today a major accuracy bottleneck preventing widespread use of OF DFT in applications.

In both KS DFT and OF DFT, it is common to reduce the computational cost by using pseudopotentials effectively replacing a real atom with a pseudoatom containing only a subset of the atom's electrons (valence electrons or more) whose electronic states match those of the real atom in the valence region. While for KS DFT, accurate (providing chemical accuracy) non-local, angular-momentum dependent, pseudization schemes are available,[29–32] OF DFT requires local pseudopotentials whose optimization poses an *accuracy* limitation. An important advantage of the pseudopotentials is the simplification of treatment of heavy atoms, whereby relativistic effects can be partially accounted for by the pseudopotential.[33] Pseudopotentials can effectively account for relativistic effects on atomic energy levels and contraction/expansion effects on the distribution of valence electrons, and obviate the need for a direct treatment of relativistic effects within DFT, i.e. solving the Dirac-Kohn-Sham equations[34]

The above-described ground state DFT provides a basis for the calculation of optical (electronic excitation) spectra with different methods, including dipole approximation often used in solid state DFT,[35] Casida-type linear response,[36,37] and real-time TD (time-dependent) DFT.[38] In the absence of orbitals, real-time TD OF-DFT can be performed based on time evolution of the electron density.[39,40] Recently, a Casida-like approach was also introduced for OF-DFT[41] based on a mapping between the real system of interacting fermions and a fictitious system of noninteracting *bosons* with corresponding single-particle wavefunctions.

## 2.2 *Brief summary of nuclear DFT*

DFT for the nuclear problem aims to describe finite nuclei and nuclear matter in terms of one-body nucleonic densities rather than the full many-nucleon wave function. In contrast to electronic DFT under the Born-Oppenheimer approximation, there is no external potential generated by clamped heavy particles.[7] A nucleus is a self-bound system, and the relevant density is therefore the intrinsic density, i.e., the density defined with respect to the nuclear center of mass.[42]

The energy is written as a functional of neutron density and proton density (energy density functional, EDF), i.e., $E_{tot}[\rho_n, \rho_p]$, which consists of kinetic energy part, nuclear interaction part, and Coulomb part,

$$E_{tot}[\rho_n, \rho_p] = E_{kin}[\rho_n, \rho_p] + E_{int}[\rho_n, \rho_p] + E_{Cou}[\rho_p] \tag{2.2.1}$$

The neutron and proton densities satisfy the particle-number constraints,

$$\int \rho_n(\boldsymbol{r})\mathrm{d}\boldsymbol{r} = N, \qquad \int \rho_p(\boldsymbol{r})\mathrm{d}\boldsymbol{r} = Z \tag{2.2.2}$$

Similar to electronic DFT, nuclear density functional theory suffers from the unknown exact form of the functional, including both kinetic and interaction parts. The unknown exact form of kinetic energy density functional necessitates the use of auxiliary single-particle orbitals, i.e., the Kohn-Sham scheme similar to electronic DFT. This elevates computational cost and restricts applications to heavy nuclear systems, such as superheavy nuclei and the inner crust of neutron stars. The lack of a known interaction functional leads to reliance on phenomenological constructions, which introduces model dependence and theoretical uncertainties.

The interaction part $E_{int}[\rho_n, \rho_p]$ represents the contribution of the effective nuclear interaction to the total energy. In practical nuclear DFT, its explicit form is not unique. It can be constructed from different classes of nuclear energy density functionals. In practical Skyrme-,[43,44] Gogny-,[45,46] or covariant[8,47] EDF calculations, this Hamiltonian contains central, effective-mass, and spin-orbit terms. For simplicity, and only as an illustrative example of a nonrelativistic EDF, we take $E_{int}[\rho_n, \rho_p]$ to be a simplified Skyrme-type functional, as used in Ref. [48]: omitting spin-orbit, Coulomb, and pairing terms, and combining the neutron and proton densities into the total nucleon density $\rho(\boldsymbol{r})$, the interaction energy reads

$$E_{\text{int}}[\rho(\boldsymbol{r})] = \int \left(\frac{3}{8}\, t_0 \rho^2 + \frac{1}{16} t_3 \rho^{2+\gamma}\right) d\boldsymbol{r} + \int \frac{1}{64}\,(9t_1 - 5t_2 - 4t_2\, x_2)(\nabla\rho)^2\, d\boldsymbol{r}, \tag{2.2.3}$$

where $t_0, t_1, t_2, t_3, x_2$, and $\gamma$ are the so-called Skyrme parameters, which need to be determined by empirical fitting to the properties of finite nuclei and the nuclear matter equation of state (the reason the parameters are carried separately in Eq. (2.2.3) and not lumped together is because they have different meanings in the full equation accounting for neutrons and protons separately, e.g. see Eqs. (10-12) in Ref. [43]). The functional derivative of Eq. (2.2.3) reads

$$\frac{\delta E_{int}[\rho]}{\delta \rho} = \frac{3}{4} t_0 \rho + \frac{2+\gamma}{16}\, t_3 \rho^{1+\gamma} - \frac{1}{32}\,(9t_1 - 5t_2 - 4t_2 x_2)\Delta\rho, \tag{2.2.4}$$

which serves as the effective potential in the Kohn-Sham equation and constitutes the gradient descent term in the orbital-free framework.

In the Kohn-Sham formulation of nuclear DFT, the densities are represented by auxiliary single-nucleon orbitals. For each nucleon species $q = n, p$, one writes schematically (we neglect partial occupancies without loss of generality)

$$\rho_q(\boldsymbol{r}) = \sum_i |\varphi_{iq}(\boldsymbol{r})|^2, \qquad q = n, p,$$

(2.2.3)

Minimization of the EDF with respect to the orbitals leads to self-consistent single-particle equations,

$$h_q[\rho_n, \rho_p]\varphi_{iq}(\boldsymbol{r}) = \varepsilon_{iq}\varphi_{iq}(\boldsymbol{r}), \qquad q = n, p$$

(2.2.4)

The effective Hamiltonian $h_q$ is generated from the functional derivatives of the EDF. In contrast to electronic DFT, it does not contain a fixed external potential from clamped nuclei. Instead, it represents a self-consistent nuclear mean field that effectively encodes the strong interaction, the proton Coulomb interaction, and other correlations included in the chosen interaction EDF. In practice, it is solved by starting from trial densities or orbitals, constructing $h_q$, solving the single-particle equation, updating the densities, and iterating until convergence, similarly as is done in electronic DFT. These cycles contribute to the CPU cost of nuclear DFT. The orbitals may be represented as basis expansions, similarly to electronic DFT. Often-used bases include spherical harmonic expansions with radial components expanded over harmonic oscillator functions[49] or Woods-Saxon basis (solutions in the Woods-Saxon potential).[50–52] Three-dimensional lattice methods can also be used.[53,54] Harmonic-oscillator bases are compact and efficient for most bound nuclei, Woods-Saxon bases are numeric bases that better capture the tail behavior in the nuclei surface and that are useful for weak bound nuclei with continuum effects, whereas coordinate-space methods are especially useful for large deformations, fission, and time-dependent dynamics.

The KS framework has been very successful for the studies of both nuclear structures and nuclear reactions. Its cost, however, is dominated by solving self-consistent single-particle equations. This becomes expensive in three-dimensional calculations, shape-constrained calculations, superheavy nuclei calculations, and also the time-dependent calculations, where the KS equations must be solved many times. Note that shape-constrained calculations are common in nuclear physics because they map the nuclear energy as a function of collective deformation coordinates, thereby enabling the study of equilibrium shapes, shape coexistence, fission barriers, and large-amplitude collective motion. This is very different from electronic DFT where three-dimensional (3D) calculations on real compounds are standard.

Orbital-free nuclear DFT returns to the density-only form of Eq. (2.2.1). The total energy is minimized directly with respect to $\rho_n$ and $\rho_p$, without introducing auxiliary orbitals. The variational condition is

$$\frac{\delta}{\delta \rho_q(\boldsymbol{r})}\left\{E_{\text{tot}}[\rho_n, \rho_p] - \sum_{q=n,p} \mu_q \int \rho_q(\boldsymbol{r}) \mathrm{d}\boldsymbol{r}\right\} = 0, \qquad q = n, p \tag{2.2.5}$$

where $\mu_q s$ are Lagrange multipliers ensuring the conservation of the total particle number. If accurate density functionals were available, this formulation would avoid the solution of large eigenvalue problems and could scale much more favorably with system size. It is very attractive for cases where computations become demanding in the Kohn-Sham scheme, such as superheavy nuclei, or nuclear matter in the inner crust of neutron stars. The central difficulty of nuclear OF-DFT is the construction of kinetic-energy functional $E_{kin}[\rho_n, \rho_p]$.

Time-dependent density functional theory (TD-DFT) provides a microscopic framework for describing nuclear excitations and collective dynamics.[55–58] Note that in the small-amplitude limit around a static EDF minimum, TD-DFT reduces to the random-phase approximation (RPA). The excitation energies and transition amplitudes can then be obtained from the linear-response equation. The Casida approach in electronic TD-DFT and the RPA method in nuclear physics[59,60] are formally analogous linear-response formulations: both arise from the small-amplitude limit of time-dependent DFT equations.

The above formalism intentionally suppresses several standard extensions. In realistic open-shell calculations, the EDF may also depend on kinetic, spin-current, current, and pairing densities. Considering pairing density will change the KS-like equations into Hartree-Fock-Bogoliubov (HFB) quasiparticle equations.[61] This is essential for quantitative nuclear surveys, but it does not alter the KS/OF distinction emphasized here: KS/HFB formulations use auxiliary orbitals or quasiparticles, whereas OF nuclear DFT seeks a density-only functional.

Finally, covariant DFT is a distinct but also widely used formulation.[8,62,63] It imposes Lorentz covariance and describes nucleons as Dirac particles moving in self-consistent scalar and vector fields, usually generated from meson-exchange or point-coupling Lagrangians. This structure naturally produces large spin-orbit splitting values and provides a compact relativistic description of nuclear saturation.[64] The formalism is therefore different: the KS-like equation becomes a Dirac equation, and the functional depends on covariant densities such as scalar density, vector density, baryon current, etc. Its time-dependent extension, usually referred to as time-dependent covariant density functional theory (TD-CDFT), applies the same covariant framework to nuclear dynamics and collective excitations. In TD-CDFT, the nucleonic Dirac spinors evolve in time under time-dependent scalar and vector fields, so that both the densities and the covariant currents become time dependent. It provides the relativistic counterpart of nonrelativistic nuclear TD-DFT/TDHF and has been used to describe giant resonances, collective vibrations, and large-amplitude nuclear dynamics.[65–70] An OF reduction of covariant DFT would thus need to preserve this scalar-vector structure rather than simply reuse the nonrelativistic kinetic-energy functional. Building OF covariant DFT remains an open question, with only very few exploratory attempts reported so far.[71]

## 2.3 *Key conceptual similarities and differences of importance to the use of ML for electronic and nuclear problems*

### 2.3.1 Conceptual similarities and differences

DFT for the electrons and nuclei bear many similarities. This is natural, as the very concept of DFT is the same, i.e., the Hohenberg-Kohn theorem is not restricted to specific particles or interactions, and therefore the governing equations used in both fields are similar. Computing energy as the functional of the density, representing the density as a sum of single particles densities that are in turn viewed as squares of single particles wavefunctions (orbitals), solving equations for the orbitals (KS equations) in a basis, updating the densities and orbitals in self-consistency cycles – all these ingredients are similar. In both electronic and nuclear DFT one uses the concept of a system of non-interacting particles where effects of interactions among particles are included in a density functional. Time evolution of the density for TD DFT can also be done similarly in electronic and nuclear DFT, also in the case of OF-DFT.[39,40,72]

A key conceptual difference is that electronic interactions are well-known (Coulombic) and serve as direct input to DFT, and the nature of correlation is well understood. In contrast, nucleons are composite particles to begin with (e.g. quark structure is ignored) so that a nucleon-nucleon interaction is already an intricate construct; nuclear ab initio theories only serve as guidance to the development of nuclear interaction and many-body theories. As a result, reliable electronic XC functionals can be derived *ab initio*, i.e. electronic DFT can be ab initio even though empirically tuned functionals are also widely used, while current nuclear DFT functionals are empirical, as exemplified by the Skyrme functional parameters in Eq. (2.2.3); many more parameters are often used than those in Eq. (2.2.3). This difference affects how ML has been used in electronic and nuclear DFT: in electronic DFT there is vibrant field of ML XC functionals aiming to build ML models of an existing (albeit exactly unknown) XC functional (see section 3.1), and beyond the construction of ML models mimicking existing approximate XC functionals, ML has been used to surpass them in accuracy or to approach the accuracy of wavefunction-based methods that serve as the reference because the underlying interactions are exactly known and are not subject to discovery. No direct equivalent of this field has emerged in nuclear DFT. ML or AI models that discover better expressions of basic interactions are still in the future. On the other hand, the way ML is used in OF DFT is quite similar in electronic and nuclear DFT, whereby one builds ML models approximating the non-interacting KEF of the KS system. Gradient expansions[73] are used in both fields as models to compare to. In electronic DFT, one also often uses terms of up to 4$^{th}$ order gradient expansion as features. A visual summary of main similarities and differences between electronic and nuclear DFT in Kohn-Sham and orbital-free paradigms as well as of opportunities for ML/AI methods to improve them that are considered below is offered in Figure 1.

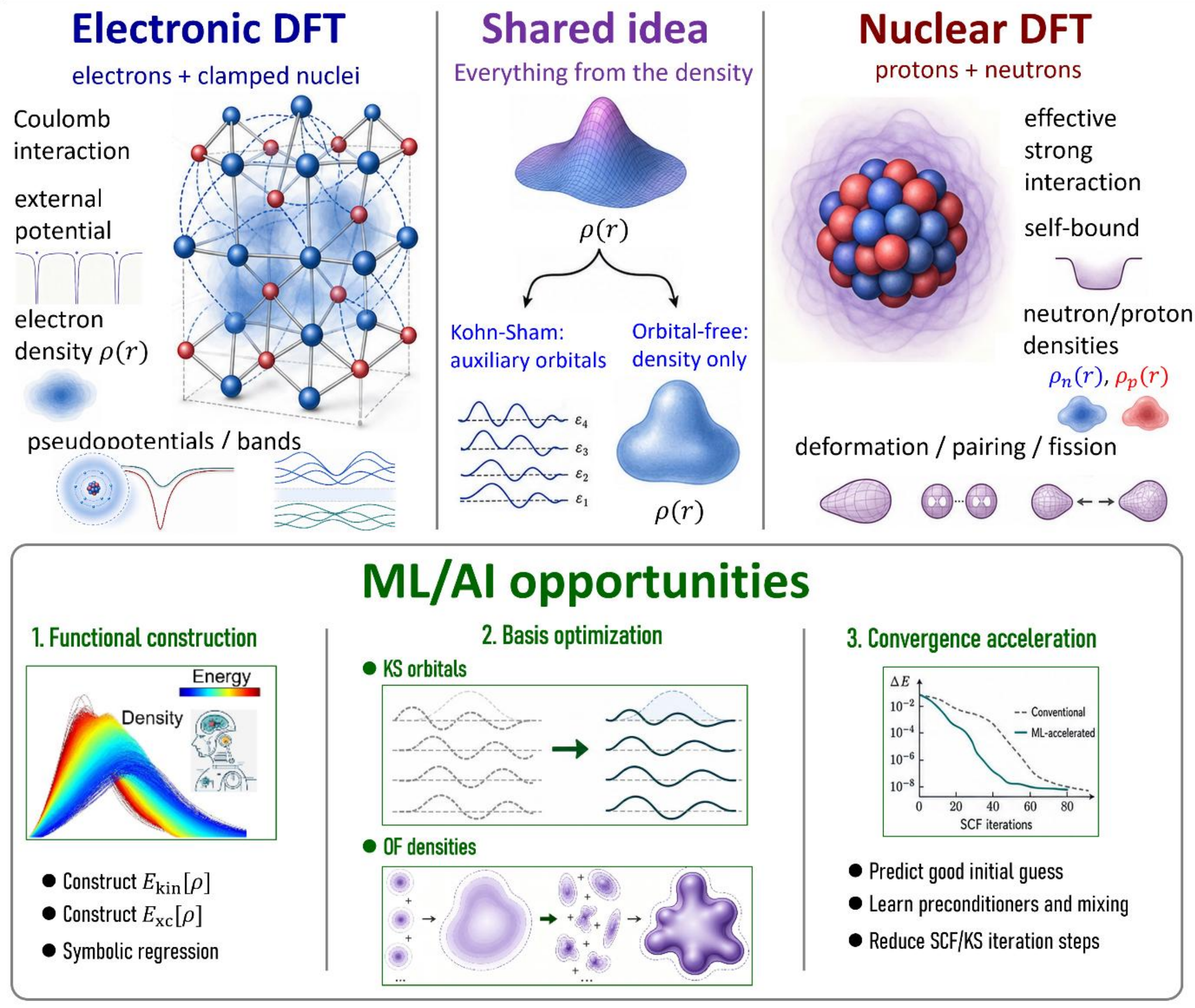

Figure 1. A visual summary of main similarities and differences between electronic and nuclear DFT in Kohn-Sham and orbital-free paradigms as well as of opportunities for ML/AI methods to improve them.

Key differences thus stem from a very different nature of particles and inter-particle interactions involved. An obvious difference is the need to consider both neutron and proton densities in nuclear DFT, corresponding to two isospin projections of the nucleon, vs. electron density in electronic DFT. The most important differences, including those that affect how ML can or should be deployed, however, stem from a different nature of interactions. In the electronic problem, electrons interact via Coulombic interactions only (and, respectively, electron density interacts with itself Coulombically). The electron density is optimized in the "external" potential due to point-wise nuclear core charges, and this potential is Coulombic and therefore divergent. This gives rise to issues not present in nuclear DFT. The divergent nature of the Coulombic potential leads to a wavefunction cusp on top of the nucleus.[74] The cusp is difficult to model: when using a plane wave basis (naturally used in most solid state calculations), an impractically large plane wave cutoff energy would be required. When using instead basis functions mimicking electronic states of the atoms (LCAO, linear combination of atomic orbitals, type bases - such bases are most often used in molecular and sometimes used in solid state calculations), Slater type functions would naturally allow modeling the cusp, but for computational expediency in the commonly used variational approach (i.e. matrix elements of Eq. (2.1.5)), Gaussian-like radial functions are used instead, increasing the basis size. Optimization of LCAO-type basis functions is an important issue, that can also be addressed by ML. Collocation allows using Slater-like functions directly, in which case basis function as well as collocation point placement optimization needs to be addressed.[75,76] Gaussian-type and Slater-type bases result in non-identity overlap matrices that may become ill-conditioned. In nuclear DFT, the situation is different. There is no fixed external potential generated by clamped point charges, and therefore no Coulombic singularity or nuclear analogue of the electronic cusp. Nuclei are self-bound systems, and the confining mean field is generated by the nucleons themselves through an effective EDF. The strong interaction is short-ranged, saturating, spin- and isospin-dependent, and includes tensor, spin-orbit, and many-body effects, which are usually encoded in Skyrme, Gogny, or covariant functionals. As a result, the main numerical challenges are not related to resolving a singular potential, but to describing self-bound densities, nuclear surfaces and tails, deformation, shell effects, pairing, and isovector dependence.[6] Harmonic-oscillator bases (used for the radial component of a spherical harmonics expansion) are compact and widely used for finite nuclei, while Woods-Saxon bases or coordinate-space grids with lattice representations are often preferable for weak binding nuclei with continuum effects, which are more suitable for large-deformed and fission configurations. While the use of a spherical harmonic expansion is similar to the electronic DFT (LCAO-type bases), in electronic DFT, LCAO-type basis functions are located on each atom (and sometimes even between atoms), while a single center is used in nuclear DFT. With a single center basis set, significant nuclear shape deformations and fission are not well represented, and one tends to use grid representation for such cases instead. In electronic DFT, harmonic oscillator basis functions have been used to generate a discrete variable representation[77] basis in which the KS equation is solved, in Ref. [78]. With a single center, the orthometric nature of harmonic and Woods-Saxon basis functions results in an identity overlap matrix, while in electronic DFT, the multicenter nature of the LCAO basis with Gaussian-like radial components requires dealing with the overlap matrix. Basis shape optimization is usually not considered in nuclear DFT, although may have potential to reduce the CPU cost of nuclear DFT calculations.

The Coulombic potential of ion cores supports core electrons that do not participate in chemical bonding and valence electrons that do. One often replaces the Coulombic potential of ion core with a smooth pseudopotential (PP) that supports only valence and sometimes some of the core electrons and is such that valence electron-derived properties (bond length, charge donation, vibrational and optical spectra, band structure of valence electrons and unoccupied bands, etc.) are the same as with the true core potential. The construction of a pseudopotential is an important optimization problem addressable by ML. The role of the pseudopotential is critical in OF-DFT where it allows avoiding dealing with rapidly changing core density, which is difficult with most KEFs. Highly accurate PPs are typically angular momentum dependent (so-called non-local) and cannot be used in OF-DFT Construction of PPs usable with OF-DFT is an important optimization problem addressable with ML that is not present in nuclear DFT; however, similar methods could be adapted for building nuclear effective potentials. Another issue not present in nuclear DFT is sampling of the Brillouin zone and calculations of the band structure (dispersion relations). In nuclear DFT, there is no direct analogue of electronic core-valence separation, pseudopotentials, Brillouin-zone sampling, or band-structure calculations. Nuclei are self-organized systems with all nucleons participating in the self-bound mean field, while the unresolved short-range and high-momentum physics of nuclear interactions are absorbed into effective EDFs. As a result, ML-addressable problems instead mainly relate to EDF optimization.

### 2.3.2 Different quantitative criteria in the two fields

Because the type of application is very different between electronic and nuclear DFT, quantitative accuracy criteria for DFT calculations are not directly comparable between the two fields. In the case of electronic DFT, one often talks about “chemical accuracy”, typically understood as energy accuracy on the order of a mHa per mol or on the order of 10 meV per mol. In practice, however, this criterion is not directly useful. DFT calculations are most often required to produce accurate structures of molecules, solids or interfaces, interaction energies, band edges and band structures (densities of states and dispersion in the Brillouin zone for periodic systems), vibrational and electronic spectra, as well as spatial distributions of electronic states (orbitals) that are important, in particular, for electronic coupling and oscillator strength calculations. Structures typically need to be accurate on the order of 1% for bond lengths and unit cell vectors and $1^0$ for angles. The required accuracy of interaction energies depends on the application, with 10 meV rarely achievable in many applications. For example, adsorption energies of molecules on catalytic surfaces or ion-host interaction energies in solid state ionics applications are rarely accurate to better than 0.1 eV. An illustration of this level of accuracy are difficulties that DFT has of accurately predicting relative phase energies.[79,80] When computing potential energy surfaces (PES), an accuracy of relative energies between different configurations on the order of 1 meV or less is needed – and is achievable – around the equilibrium geometry for accurate vibrational spectra; an accuracy on the order of 10-100 meV is more common for reactive PES over wide energy and geometry ranges. Accuracy of band edges is on the order of 0.1 eV with hybrid functionals, while with GGA (generalized gradient approximation) functionals most often used for extended systems, the band gap is underestimated by a whopping 1/3 without preventing the use of such calculations for relative comparison of band structures, e.g. for band alignment in devices. Errors in fundamental frequencies in vibrational spectra, and therefore in the curvature of the PES at equilibrium geometry, are typically on the order of a few %, and errors in electronic excitation energies are on the order of 0.1 eV only when optimal functionals are used for the type of excitation considered, and much higher otherwise. The above accuracy ballparks are for standard, not ML-enhanced, DFT, and they are indicative of what ML-enhanced methods need to beat or to achieve while offering other advantages such as better speed of calculation or better accuracy for parts of chemical space than non-ML-enhanced DFT.

In nuclear DFT, the accuracy criteria are different and observable-dependent. Binding energies, separation energies, charge radii, deformations, fission barriers, collective excitations, and nuclear-matter properties are all relevant. Global EDF optimizations typically tolerate errors of around 2 MeV in total binding energies and about 0.03 fm in charge radii.[51] Shell gaps, fission barriers, and single-particle splittings often demand sub-MeV to about several MeV accuracy.[81] Collective excitation energies are commonly accurate only at the level of several hundred keV to a few MeV.[82] Nuclear potential energy surfaces (energy change as a function of nuclear shape) need to be accurate on the order of $10^{-1}$ MeV. Thus, there is no single nuclear analogue of "chemical accuracy." For ML-enhanced nuclear DFT, the goal is instead to improve the accuracy of the target observables, reduce the computational cost of large-scale EDF calculations, provide reliable uncertainty estimates, and ensure a controlled description of the relevant physical effects for the problem under consideration. While, depending on the purpose of the calculation, different measures of accuracy are used within each electronic and nuclear DFT, overall one can say that electronic DFT has more stringent requirements on relative energy accuracy in most applications: interaction energies that need to be accurate on the order of $10^{-3}$-$10^{-1}$ eV are computed from differences in total energies that are often on the order of $10^{3}$-$10^{6}$ eV. In nuclear DFT, relative accuracy on the order of $10^{-1}$-$10^{0}$ MeV is needed while total binding energies are on the order of $10^{3}$ MeV. The relative energies, e.g., energy differences between different chemical compositions, isomers, phases or between different geometries in electronic DFT, and accurate relative binding energies between different nuclei or between different shapes of a nucleus in nuclear DFT, are usually the quantity of interest rather than absolute total energies. High relative energy accuracy can be achieved either with models embedding error cancellation - which is an important advantage of physics-informed analytic models - or with very high absolute energy accuracy without error cancellation. As major ML methods by themselves do not incorporate error cancellation, the relatively milder relative error requirements in in ground-state energies of nuclear DFT applications create an opportunity for ML-enhanced nuclear DFT to be adopted for applied calculations sooner than electronic DFT for which the higher relative accuracy requirements may be harder to achieve. This concerns especially OF-DFT where a ML KEF needs to accurately model a significant fraction of the total energy as opposed to, e.g. the XC component only. Alternatively, development of physically-motivated ML models that can embed error cancellation is desired (as opposed to using off-the-shelf ML methods as is).

# 3 Key computational and accuracy bottlenecks that can be addressed with ML and AI methods

We now consider applications of ML and ultimately AI methods to improve the accuracy and or CPU cost of DFT calculations, by applying ML/AI methods to compute energy density functionals, electron or nucleon density, to improve basis completeness or accelerate convergence. What follows is not an overview of ML-enhanced DFT, but rather a comparative assessment of promising ML/AI methods for different kinds of DFT: electron and nuclear, Kohn-Sham and orbital-free. We note that reviews of ML uses for enhancing particular flavors of DFT have been appearing, mostly for electronic DFT, and the reader can refer to them for wider context.[83–89] In particular, relatively recent reviews of ML XC functionals[88] and ML-enhanced electronic OF-DFT[83] can be recommended. There are no dedicated reviews of ML-enhanced nuclear DFT; for a review of ML in nuclear physics, the reader can refer to Ref. [90].

### 3.1 Density functionals and related quantities

In electronic DFT, ML functionals are typically built to represent the exchange correlation functional[91–105] (XC energy or energy density, often as enhancement factors to existing functionals; or XC or KS potential[106–112]), the kinetic energy or KED functional[113–126] (directly, via its functional derivative i.e. Pauli potential $v_P = \frac{\delta E_{kin}}{\delta \rho(\boldsymbol{r})}$,[125,127–129] or related quantities[123]), or the universal functional $F[\rho(\boldsymbol{r})]$ (Eq. (2.1.1)).[130] The field of ML functionals exists because existing formulaic analytic functionals are insufficient. There exist a large number (hundreds[131]) of XC functionals finding use in applied electronic DFT calculations, from multiple GGA functionals used in most solid-state simulations to multiple global and range-separated hybrid functionals used in most molecular simulations to meta-GGA functionals. This diversity reflects the fact that there is no single functional providing simultaneously practically sufficient accuracy of structures, band structures, and optical properties for most materials, but different types of functionals do provide practically relevant accuracy of specific properties. Even then, the accuracy is often semi-quantitative, so a quest for ML XC functionals can be justified on both the accuracy and chemical space coverage grounds. Another use of ML XC which may become more important going forward with the advent of more accurate KEFs is using an ML XC to mimic hybrid (i.e. orbital-dependent) XC functionals in an orbital-independent way.[95,102]

The issue of accuracy and chemical space coverage is much more severe in OF-DFT, as available formulaic functionals only reliably provide near-KS DFT accuracy for simple metals.[132–138] While XC functionals need to reproduce an important but relatively small part of total energy, KEFs must accurately reproduce a much more significant part of total energy. As ML methods typically do not exhibit error cancellation behavior of physically-motivated models (when computing interaction energies), this results in accuracy requirements that are much higher than those for other ML applications (such as materials informatics or interatomic potentials). For example, Ref. [119] demonstrated that correlation coefficients between predicted and ground truth values as high as 0.9999 for the ML functional to correctly (semi-quantitatively) reproduce energy-volume curves of several hundred compounds.

For nuclear DFT, the field of ML functionals has so far focused mainly on KEFs.[139–142] Traditional nuclear orbital-free approaches have mainly been based on semiclassical Thomas-Fermi or extended Thomas-Fermi approximations, which can describe average bulk properties but do not naturally reproduce quantum shell effects and deformation effects.[4,5,143] This is a particularly severe issue in nuclei, because shell structure, spin-orbit effects, and deformation are central to binding energies, radii, and potential-energy surfaces. To incorporate the shell and deformation effects, additional corrections must be applied,[144,145] which would however involve the auxiliary one-body orbits again and are no longer orbital-free DFTs. Note that a recent study[48] shows that nuclear shell effects can be incorporated by introducing a nonlocal kinetic energy density functional, whose nucleon localization functions reproduce the shell-related fluctuations obtained from Kohn-Sham calculations.

The state of the art of ML KEF in nuclear OF DFT allows reproducing nuclear potential energy surfaces, i.e., total energy as a functional of nuclei deformation, as demonstrated in Figure 2 reproduced from Ref. [142]. This issue is conceptually similar to the requirement for the KEF in electronic DFT to reproduce the shape of the potential energy surface, an issue that was problematic for a long time[78,146] and is now being effectively resolved with ML KEFs for many molecules and solids.[114,118,119,126,147]

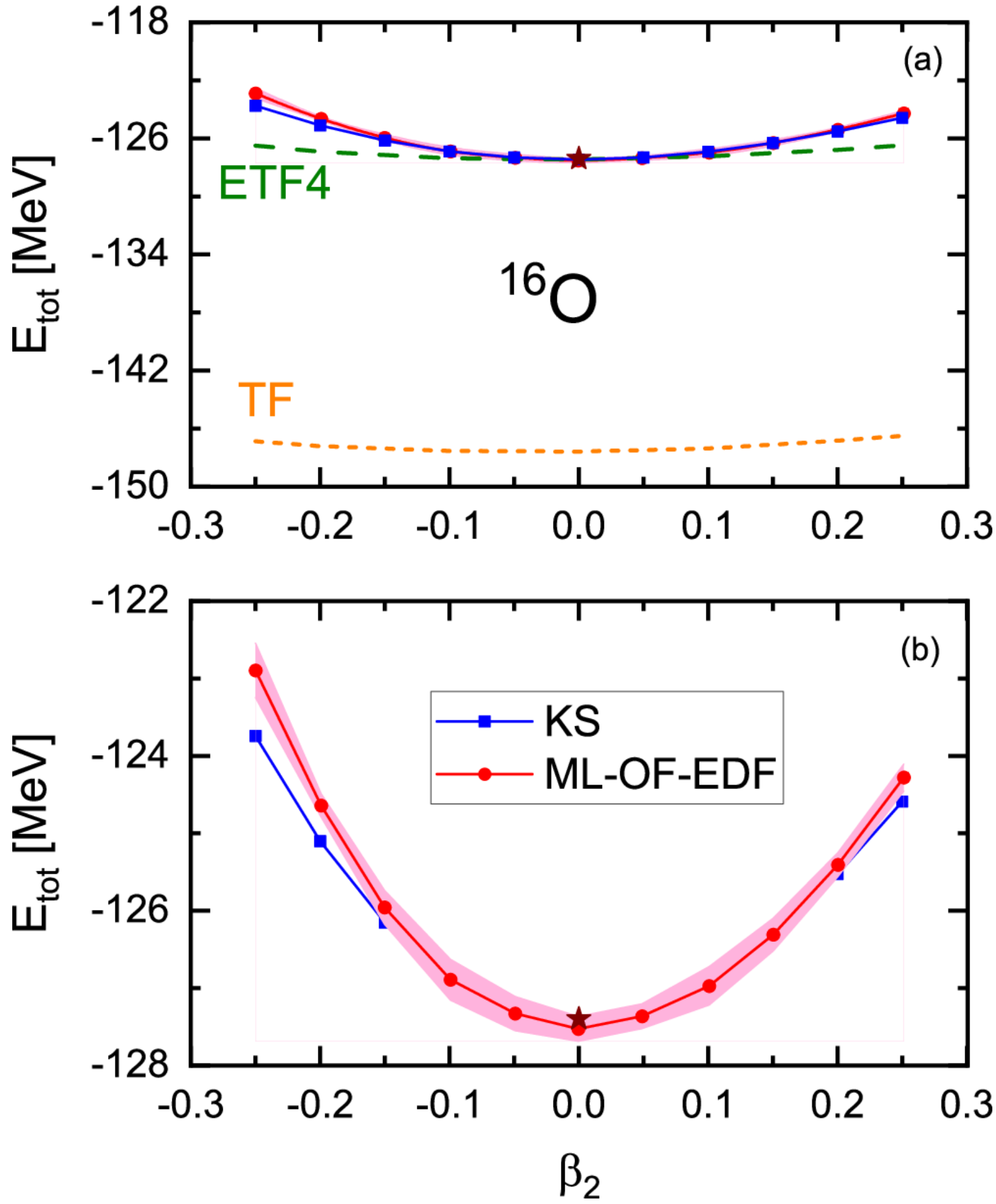

Figure 2. Potential energy curves of $^{16}O$ nucleus computed with KS DFT and with OF DFT using an ML functional in Ref. [142], as a function of the quadrupole deformation parameter $\beta_2$. Results from Thomas–Fermi and extended Thomas–Fermi (up to the 4th

order gradient expansion) models are also provided for comparison. Reproduced from Ref. [142] under Creative Commons BY-NC-ND 4.0 license. (c) 2025 The Authors.

ML functionals could be categorized as pure density functionals that use only density functionals-dependent features, and functionals that also use features encoding the structure. In the case of electronic DFT such features would encode chemical composition and structure, in the case of nuclear DFT, this would correspond to using information about the nuclear shape; use of $N$ and $Z$ as "composition" features[148] would not make an ML functional "impure" as those themselves are determined by the (integral of the respective nucleons') density. With ML functionals that depend on the density as well as on structure, the resulting calculation can be viewed as a hybrid between DFT and an ML interatomic potential (force field), as force fields, both ML and formulaic, predict energy in function of structure and may also carry a model of electron density.[149,150] As density featurization in electronic DFT is often done using basis sets that include functions (such as Gaussian functions) localized on atoms, most ML functionals (including practically all recently proposed XC functionals) include structural information as their inputs and are thus not pure density functionals.

Although the first known XC functional used a single hidden later NN with only 8 neurons,[112] and the second known ML XC functional used a single hidden later NN with only 3 neurons,[99] recently, often complex large NNs, came to dominate ML functionals in electronic DFT. A Behler–Parrinello type neural network - an NN architecture originally developed for interatomic potentials[151,152] - was used to fit the XC functional in Ref. [95] and showed better results than common functionals such as PBE, SCAN, ωB97M-V for small polyatomic molecules. A transformer NN was used in Ref. [91] to learn nonlocal XC functionals mimicking $r^2$SCAN and TPSS meta-GGA functionals and allowing accurate automatic differentiation.[153] A fully connected three- hidden layer, 300 neurons NN was used in Ref. [96] to fit an enhancement factor to SCAN functional, achieving an accuracy somewhat higher than SCAN for several materials (including C, Si, InP, MgO, and MgS). A fully connected three-hidden layer, 300 neurons NN was used in Ref. [97] for a meta-GGA functional. A similar size NN was used in Ref. [154]. 400 neurons in 2 hidden layers were used in Ref. [101]. Ref. [155] used 1200 neurons in 4 hidden layers. A fully connected NN with 10 layers and 512 or 1024 neurons per layer was used in Ref. [105], which offered a generic software tool for fitting ML functionals in the form of a weighted sum of energy densities (of other known functionals), where the (space-dependents) weights are machine-learned. The DM21[94] functional has a similar form. A CNN was used to ML the XC potential in Ref. [111], slightly outperforming B3LYP for $H_2$, $HeH^+$ and He2.

Exchange correlation potential was machine-learned in Ref. [106] with a relatively large NN even in 1D (2 hidden layers with 300 nodes) with ReLU . While the use of ReLU neurons (which has been used in several works) is attractive from the perspective of computational expedience, we would warn about its discontinuous nature, which leads to larger NN and derivative discontinuities. Equivariant GNNs was used in Ref. [103] to make an XC functional approaching CCSD(T) accuracy on MD17 molecular dataset.[156] Density matrix was machine-learned in Ref. [157] using equivariant multihead attention graph neural network for $H_2O$ and $CH_4$, diamond, and graphene. This type of NN was also used in Ref. [158] to machine-learn DFT Hamiltonian matrix (testing on $H_2O$, $MoS_2$, and Na); accuracy close to "traditional" DFT was obtained for the materials considered in these works. *An equivariant GNN can have millions of parameters even for medium-sized molecules.* Physically constrained NNs are often used, for example, in Refs. [104,105] to ML the XC functional and in Ref. [107] to ML the Hamiltonian and its derivatives used to compute the KS potential (in 1D) for TD DFT. Recently, the utility of quantum NN (QNN)[159] (a type of deep NN suitable for quantum computing) for XC functional representation has been demonstrated not only on model sub-dimensional systems but also on real small molecules ($H_2$, $H_4$, $H_2$-$H_2$) achieving mHa accuracy.[92] A kernel regression was used in Ref. [160] with SOAP[161]-based density encoding to ML the correlation energy (with CCSD(T) correlation as target) of water and hydrocarbon molecules. Note that the use of a second-order polynomial kernel is equivalent to a second-order polynomial models, suggesting that relative simplicity of underlying functional dependence (between the chosen target and features). An RBF kernel with SOAP-based features was used to ML CCSD(T) correlation in Ref. [162]. Gradient boosted decision trees (GBDT)[163] has been explored in a limited way but could obviate disadvantages of large NNs. For example, XGBoost[164] was used in Ref. [100] to improve on the PBE functional; the authors explicitly stated that XGBoost outperformed NNs in their work and highlighted the unwieldy nature of large NNs.

As mentioned in section 2.3, no direct equivalent of the field of ML XC functionals has emerged in nuclear DFT, in particular, due to the postulated nature of elementary interactions (as opposed to the exactly known Coulombic interactions in electronic DFT). However, this creates an opportunity for the use of ML or AI methods to discover better expressions of basic or effective nuclear interactions in the future.

An important difference between the KS and OF approaches is the different relative importance of the computational cost of calling the ML model of the functional. In the KS paradigm, a major determinant of the CPU cost and its nonlinear scaling with system size is the diagonalization of the Hamiltonian matrix. The overhead of the computational cost of calling an ML model of the XC functional is *expected* to be not significant. This changes in the OF paradigm, where most of the computational cost comes from evaluating the energy functional. The recent proliferation of large (excessively large, in the authors' view) NN models indicates a direction that partially defeats the purpose. This is compounded by the fact that ML models tend to have worse convergence as their errors are more corrugated that those of analytic formulas (which is itself related to complex architectures), increasing the risk of convergence problems in both self-consistency cycles and geometry optimization, as was recently convincingly highlighted in in Ref. [165]. As an example, we reproduce in Figure 5 computational time tests performed in Ref. [165] using of DFT with the analytic hybrid PBE0 and machine-learned DM21[94] functionals as well as CCSD(T) (coupled cluster with singles, doubles, and perturbative triples, considered as the "golden standard" wavefunction-based ab initio method). The comparison is to a hybrid functional as DM21 depends on local HF exchange energy density, making it a hybrid ML functional. While PBE0 easily beats CCSD(T) in computational efficiency already for small systems, DM21 has a higher cost than CCSD(T) even in the KS regime! DM21 demonstrates that as of today ML functionals can compete with the best analytic functional and beat them on specific metrics, e.g. DM21 addressed the problem of delocalization error in substance. As any ML models, they require training data (e.g. more than 4000 molecular structures with main-group elements in the case of DM21) and suffer from extrapolation errors / overfitting; this is directly related to the complexity of the architecture. Ref. [165] studied how DM21 extrapolates to transition metal chemistry (TMC) and concluded that DM21 achieves comparable or occasionally superior accuracy to B3LYP but struggles with self-consistent field convergence. On the other hand, Ref. [103] reported more favorable CPU time comparison results of their GNN-based functional with several traditional functionals including LDA (local density approximation), hybrid, and meta-GGA functionals, whereby the NN functionals were faster than B3LYP or ωB97X . Interestingly, the CPU time comparison in Ref. [103] also included the NN functional of Ref. [104] which used a relatively small NN with three hidden layers with 16 neurons each, yet the timing performance of the two was by and large similar. Note that efficient GPU (graphics processing unit)- and TPU (tensor processing units)[166]-amendable implementations of complex NN architectures play an important role here; the underlying math of GNNs remains compute-heavy.

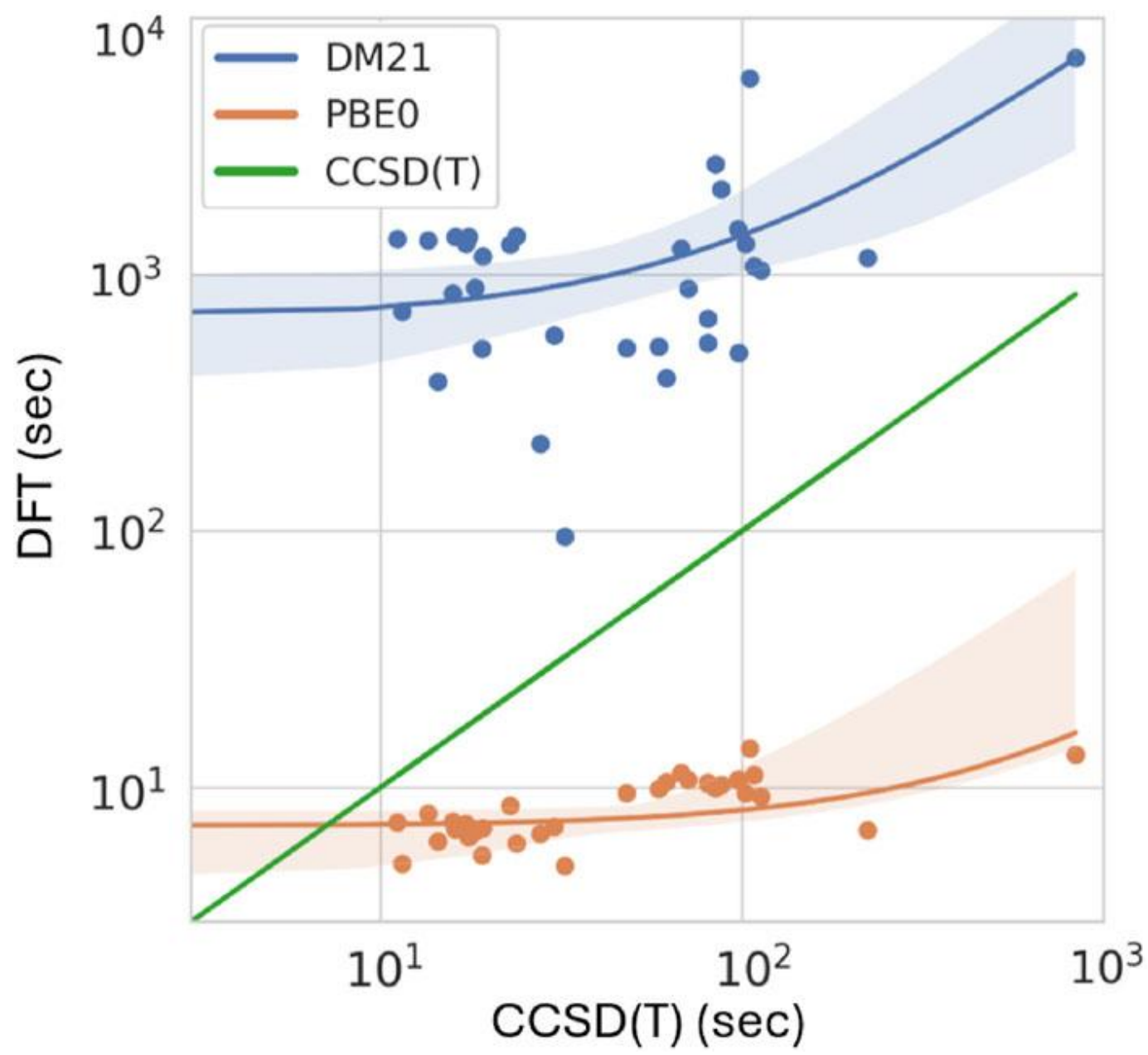

Figure 3. Computational time tests of DFT with PBE0 and DM21 functionals and CCSD(T). The horizontal axis shows the computation time for a single SCF cycle for two-atom molecules made of elements from H to Ca (neutral form and cations) using the cc-pVTZ basis set and an energy tolerance of $10^{-9}$ a.u. using the CCSD(T), while the vertical axis shows the DFT computation time of these systems with DM21 and PBE0 functionals.

A similar issue exists with KEFs, where the use of relatively large convolutional NNs, GNNs, graphomers etc. has been proliferating[121,126,129,144,147,167–169] - see Ref. [83] for a more comprehensive overview of ML KEFs for electronic OF DFT. While the short-range nature of the functional[170,171] (justifiable by the properties of one-electron reduced density matrix[172,173]) means that the scaling can still be linear with system size for large enough systems, the use of oversized ML models increases the CPU cost prefactor. “Lighter” models are therefore important for OF-DFT. Besides formulas obtained with symbolic regression (see below), one can stress the importance of using lighter ML methods. Ref. [116] demonstrated that an NN need not be large if fitted with a powerful optimizer (such as Levenberg-Marquard) and with physically motivated features such as terms of the gradient expansion: they any needed an 80-node (20 per hidden later) NN to reproduce KS KED of Si, Li, Al, and Mg simultaneously; for individual materials single hidden layer NNs with only a few neurons were sufficient. Non-NN methods such as kernel regressions with which the field of ML KEF largely started first with 1D models[78,174] and which have been shown to be promising for real materials[117,119,120] can also result in lighter models. GBDTs (gradient-boosted decision trees) also deserve to be explored for KEFs, especially as available results for ML XC functionals are promising.[100] Both kernel-based and NN-based methods have been applied in nuclear DFT, yet their computational costs have not been compared.[139,140]

### 3.1.1 Role of sub-D models in electronic and nuclear DFT

Sub-dimensional (one- or two-dimensional) ML functionals carry different importance in electronic and nuclear DFT. In electronic DFT, sub-dimensional models served to explore the idea of constructing ML functionals and to test different ML methods and featurization methods and to explore way to obtain stable functional derivatives with ML functionals, which is important for self-consistent optimization.[78,84,106,107,127,130,169,174–179] They are of very limited use to the simulations of real molecules, solids, and interfaces, where an ML model must deal with realistic 3D electron densities. Some ideas explored on sub-dimensional models, such as density featurization using grid values or related PCA approaches,[169,174] are difficult to infeasible in 3D simulations on real materials. Such sub-dimensional models also used simplified analytic synthetic interactions potentials (rather than real atoms'). However, we note that pseudo-2D models have been quite useful for ultra-large-scale simulations, for example DFT based quasicontinuum simulations up to μm scale,[180–182] and sub-dimensional ML functionals could be useful for such simulations. This direction is still unexplored.

In nuclear DFT, however, sub-dimensional models could have a more direct practical role. Many realistic nuclear-structure calculations are performed under symmetry restrictions. For spherical or near-spherical nuclei, the local densities reduce essentially to radial functions, so that a one-dimensional representation can already describe a physically relevant class of systems. For axially deformed nuclei, the densities are commonly represented on a two-dimensional grid. Such reductions are not merely pedagogical simplifications, but are routinely used in real calculation such as large-scale nuclear mass-table calculations.[51] Therefore, ML functionals trained and tested in one or two dimensions could be connected to bona fide applications rather than only to toy model systems.

### 3.1.2 Beyond traditional ML

Beyond ML, techniques that could be called AI are still little used to build functionals. Symbolic regression has been used in a limited way and presents significant potential going forward. Automatically constructed formulas have the advantage of CPU cost, easy software integration into DFT codes, and interpretability / insight that can be used for further theory development. Symbolic regression has been used for XC functionals. In Ref. [93] symbolic functional evolutionary search was used to recover ωB97M-V functional as well as to develop a new functional that outperformed it on most molecules in Main Group Chemistry Database (MGCDB84).[131] The method evolves populations of formulas made from building blocks of existing functionals, elementary terms including electron densities and gradients, and functional forms of enhancement factors were learned. To illustrate how symbolic functional evolution works, we reproduce the workflow of the method in Figure 4. Genetic evolution had also been used in Ref. [183], whereby the genome included pieces of known GGA, meta-GGA and hybrid XC functionals; an improved version of the B3PW91 functional was discovered.

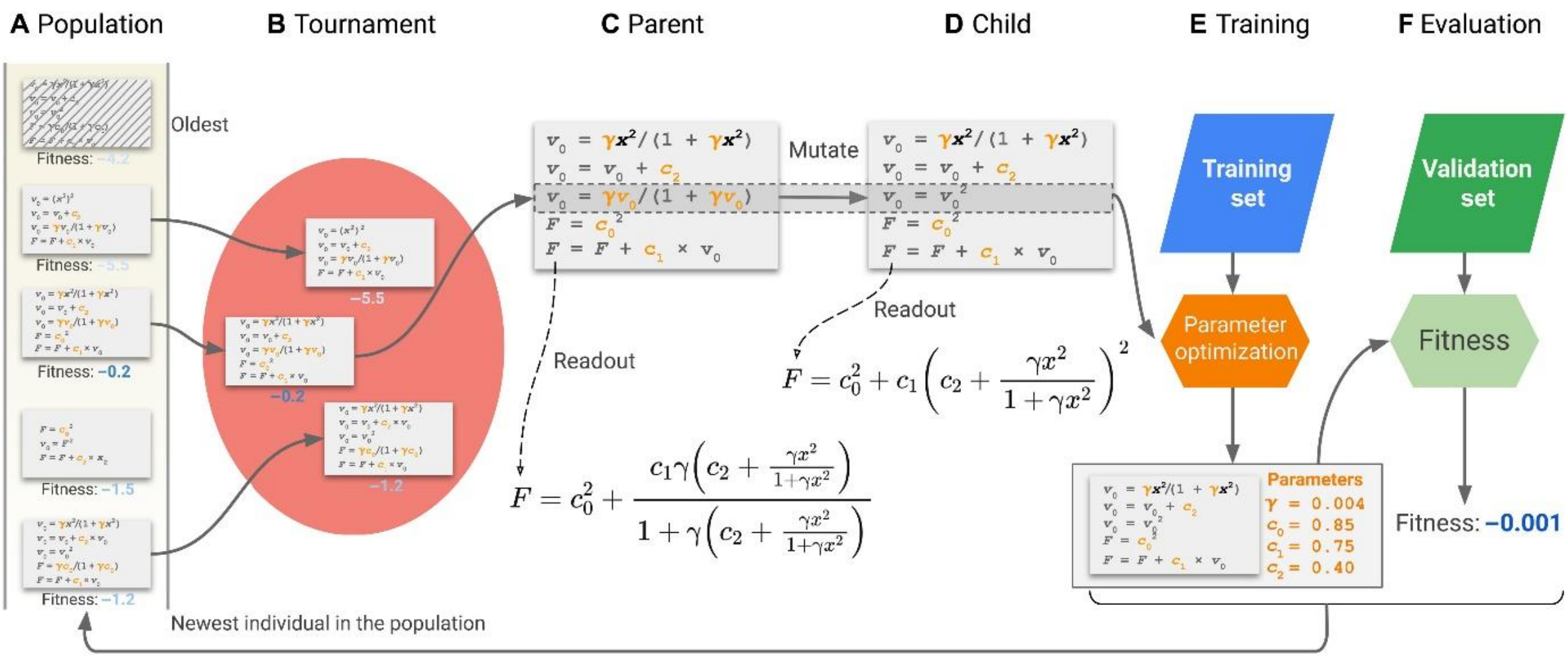

Figure 4. Workflow of the symbolic functional evolution method use to discover XC functionals in Ref. [93]. (A) A population of symbolic density functionals is iteratively evolved using an evolution algorithm. Each individual in the population represents the enhancement factors in a density functional. The performance of a density functional is characterized by its fitness, defined as the negative of validation energy error. (B) In each iteration, a tournament selection is performed on a subset of individuals sampled from the population. (C) The highest-fitness individual in the tournament is selected to be the parent. (D) Then, a child functional is generated by randomly picking one of the enhancement factors and mutating one of its instructions. (E) The child functional is then trained on the training set using the covariance matrix adaptation evolution strategy (CMA-ES) method to determine its parameters. (F) Last, the fitness of child functional is evaluated on the validation set and then added to the population. After the population size exceeds a limit, the oldest individual in the population is removed. For simplicity of visualization, a density functional is represented with only one enhancement factor. The general form of density functional considered in this work contains three enhancement factors, and in each evolution step, one enhancement factor is randomly chosen for mutation. Reproduced from Ref. [93] under Creative Commons Attribution License 4.0 (CC BY). (c) 2022 The Authors.

Several papers appeared recently on symbolic regression for electronic OF DFT. Ref. [176] used genetic programming to learn the non-interacting KEF for a 1D system. The method was able to discover the von Weizsäcker functional for single-electron systems and approached the Thomas–Fermi limit for multi-electron systems. Symbolic regression-derived functionals were also stable with respect to the functional derivative. Ref. [184] use a physics-inspired symbolic regression method to rediscover the TF functional and was used to build a γTFλvW functional (where γ and λ are parameters) tested on H, $He^+$, $Li^{2+}$, He, Be, Ne, Mg, Ar, $H_2$, HF, $N_2$, CO, $O_2$, and $F_2$. As of now, the only example of symbolic regression of an electron density functional for real materials beyond TF – vW models is Ref [118]. That work employed the GPR-NN method[185] which is effectively a hybrid between NN and GPR and allows analyzing the kind of functional dependence of the target function on the features.[186,187] The method is thus easily amenable to symbolic regression, and was used in Ref [118] to build the first ML-guided KEF formula for real materials. The features were made from the terms of the 4$^{th}$ order gradient expansion[73] as well as the product of electron density and Kohn-Sham effective potential, $\rho v_{eff}$, which is related to the kinetic potential[125] and is responsible for a significant fraction of the variance of kinetic energy density.[117] Excluding , $\rho v_{eff}$, it was possible to test the approach on the recovery of the 4$^{th}$ order gradient expansion, as the model in that case return a gradient expansion-like expression with coefficients fitted to a particular dataset (which included 433 unary, binary, and ternary materials containing Li, Al, Mg, Si, As, Ga, Sb, Na, Sn, and P[119]) and with some nonlinearity applied to the terms of the expansion. The ML model returned physically meaningful relative magnitudes of the terms related to the terms of the gradient expansion, demonstrating interpretability.

Symbolic regression still has not been used in nuclear DFT where it offers a promising direction of research. Its use in nuclear DFT would nevertheless differ in several respects from applications in electronic DFT. In electronic DFT, symbolic regression of the functionals can often be guided by the well-defined Coulomb interaction. In nuclear DFT, the energy density functional is more complicated and must effectively encode correlations associated with the strong interaction and with many-body nuclear dynamics. The relevant variables are also richer: in addition to the total nucleon density, one may need to distinguish neutron and proton densities, or equivalently isoscalar and isovector densities, and realistic functionals may also depend on kinetic, spin-current, current, and pairing densities. Symbolic regression in nuclear DFT would have to be constrained not only by finite-nucleus properties, such as binding energies, charge radii, deformations, spin-orbit splittings, pairing effects, surface properties, etc., but also by nuclear matter properties, such as the symmetry energy, equation of state, etc.

### *3.2 Orbital and density representation*

#### 3.2.1 Basis sets for the solution of the Kohn-Sham equation

We consider here basis sets used to solve the KS equation, i.e. to convert the KS equation into a matrix equation. Machine learning of basis functions is done to produce a sufficiently complete basis of a small size, thereby reducing the size of the matrices in Eq. (2.1.5). We mentioned above similarities and differences between the basis types used in electronic and nuclear DFT. In particular, LCAO type bases used in the electronic DFT are multi-center, non-direct product type, and result in a non-identity overlap matrix which may be badly conditioned especially for highly complete bases. The single center nature of the bases often used in nuclear DFT which are made from spherical harmonics and orthonormal radial functions results in a direct product basis with an identity overlap matrix. As a key advantage of ML is representation of a multivariate function in, effectively, a parameterized non-direct product basis made e.g. from neurons of a NN or from kernel functions of a kernel regression (these two methods being major regression type ML methods), this distinction is conceptually important. ML-optimization of radial functions or of the overall shape (including angular components) does not conceptually change the non-direct-product, non-orthonormal nature of the basis in electronic DFT while it would in nuclear DFT and could introduce additional difficulties in handling non-orthonormality of the basis. The number of basis functions is on the order of dozens to hundreds per center in both electronic and nuclear DFT. As it scales with the number of atoms, and as the CPU cost of solving the KS equation is cubic in basis size, the basis size is a limiting factor in electronic DFT (limiting feasible system sizes), and basis optimization – either manual or ML-based – is impactful.

Several works offered approaches to optimize, with the help of ML, basis shapes in function of chemical environment for electronic DFT. For example, in Ref. [188], LCAO-type basis functions were machine-learned using GPR and rotationally invariant descriptors of chemical environment that were eigenvalue of the overlap matrix constructed from Gaussians located on the target atom and its surrounding atoms. Interestingly, the authors proposed to machine-learn not basis parameters directly but parameters of a polarization potential (made from Gaussian functions) in an auxiliary atomic Hamiltonian, whose eigenstates were used to make basis functions. As a result, a more compact basis set could be used for the same accuracy and a more than 50-fold speedup of calculations vs. standard bases (from single- to triple-zeta polarized) was reported on calculations of liquid water. Frames of ab initio MD on a 64-molecule model of liquid water were used to acquire training data. A similar approach was used in Ref. [189] with LASSO regression that was used to fit potential parameters to training set of 118655 data points produced by coupled cluster calculations of molecules containing H, B, C, N, O, F, Si, P, S, and Cl. Thus optimized double-zeta bases could produce molecular properties (interaction energies and conformational energies, reaction energies, barrier heights, and bond separation energies) with accuracy close to complete basis set wavefunction (i.e. more accurate than DFT) methods (i.e. this work had also elements of delta and transfer learning). In Ref. [190], ML was used to optimize parameters of muffin tin orbitals (MTO)[191] – a basis type that is LCAO-like in the core region and plane wave-like in the valence region, allowing, in particular, efficient full-potential (as opposed to pseudopotential) calculations. The authors used covariance matrix adaptation evolution strategy (CMA-ES) algorithm. Parameters (such as atomic sphere radii) were thus fitted to minimize the discrepancy with plane wave calculations. In this way, improved band structures for $CrTe_2$, Si, GaAs, and $CrI_3$ were obtained. How extensive basis set optimization may need to be shows the example of BigDFT code, where, starting from an LCAO-like basis, basis functions are expanded in wavelets and optimized to minimize total energy, effectively obviating a key downside of LCAO-type bases.[28,192] An example of the optimized function starting from a *p*-type Si-centered function, for a hydrogen-terminated Si nanoparticle, is shown in Figure 5. Predicting such optimized functions with ML would significantly accelerate large-scale DFT calculations. Note that works that ML the DFT Hamiltonian matrices,[193,194] do so in a particular basis, typically LCAO type. In Ref. [194], to help the use of training data computed with plane wave codes that are currently much more widely used for solid state calculations and are used to generate major DFT data dabases,[195,196] DFT Hamiltonian matrix elements in LCAO type basis were predicted from those in a plane wave basis.

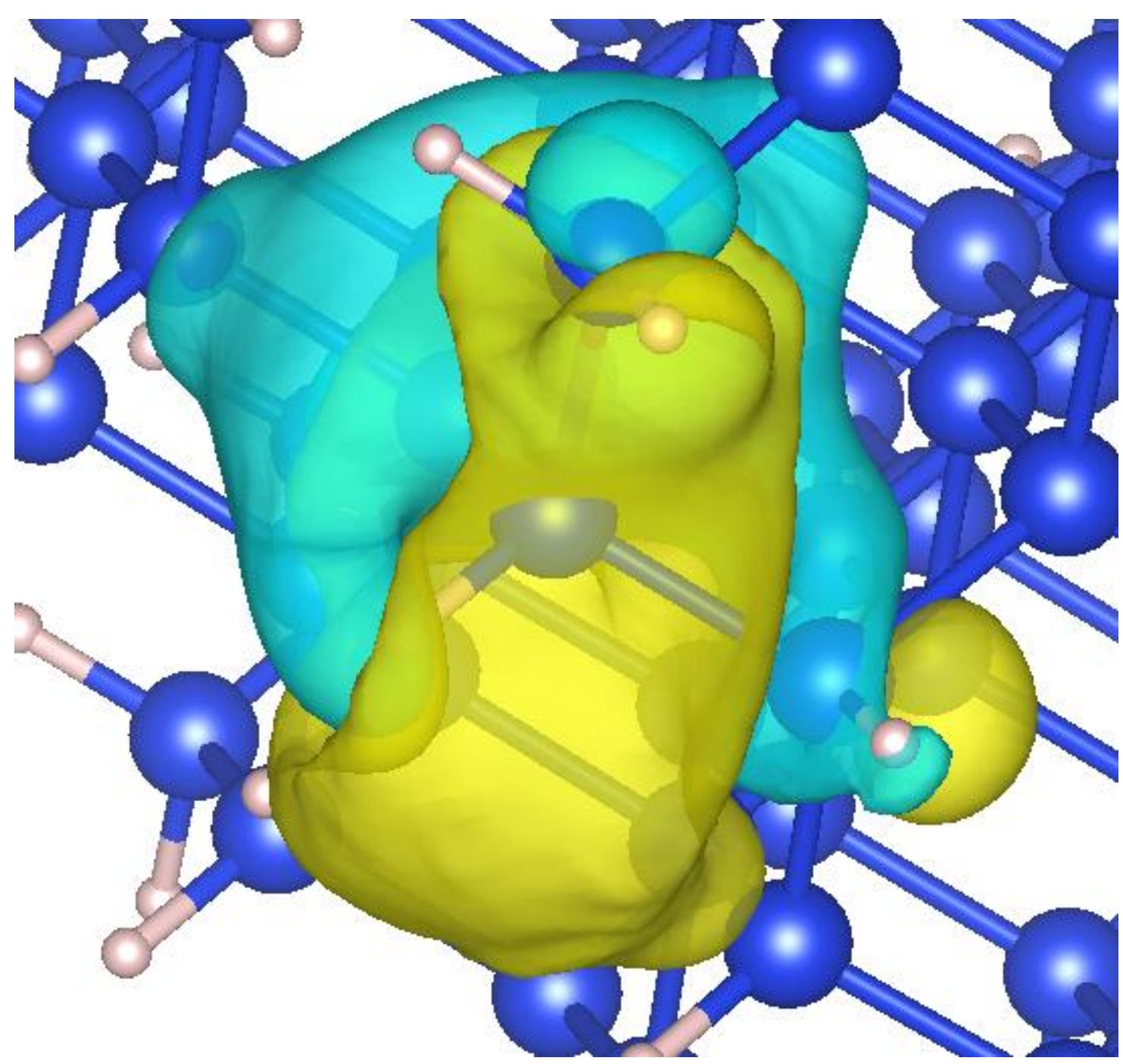

Figure 5. Example of a basis function optimized in BigDFT[192] to minimum energy, starting from a *p*-type Si-centered function, for a hydrogen-terminated Si nanoparticle.

The issue of the wavefunction cusp in the electronic problem that complicates integration in the context of variational calculations can be obviated by using collocation.[76,197] In Ref. [75], it was shown that ML can be used to optimize collocation point placement. In that work, the basis was of spherical harmonic type with the parameters of the radial component optimized manually. Their optimization with ML would offer further improvement of basis quality. Collocation has not yet been explored for nuclear DFT. One its advantage is the possibility of using non-integrable basis functions[198] which could offer additional optimization opportunities.

Basis size is generally less of a limiting factor in nuclear DFT than in electronic DFT, especially for near-spherical or moderately deformed stable nuclei, where harmonic-oscillator bases are usually compact and efficient.[49] However, the choice and optimization of the basis becomes important when the basis is poorly adapted to the relevant density distribution. This occurs, for example, in weakly bound nuclei close to the drip lines, where the density may have an extended asymptotic tail that is inefficiently represented by the Gaussian falloff of a harmonic-oscillator basis, and in large-amplitude collective motion such as fission, where strongly elongated, necked, or fragment-like configurations are not naturally described by a single-center basis.[199–201] Triaxial and reflection-asymmetric shapes, superheavy nuclei, multidimensional potential-energy-surface calculations, and large-scale mass-table calculations can further amplify the cost and truncation errors associated with basis size because many self-consistent calculations have to be repeated.[51,52,202–204] Existing non-ML developments already demonstrate the practical value of adapting the basis to the nuclear problem. Deformed harmonic-oscillator bases are widely used for the calculations of deformed nuclei;[202,205,206] transformed harmonic-oscillator bases were introduced to improve the description of weakly bound systems and their asymptotic densities;[203,207] Woods-Saxon basis and its optimization provide a more realistic representation of surface, asymptotic, and continuum effects;[208,209] two-center harmonic-oscillator bases are useful for fission and fusion configurations in which a single-center basis is inefficient.[201,210] Recent studies of basis truncation and oscillator-frequency optimization in covariant DFT also show that basis parameters can have a measurable effect on calculated binding energies and related observables.[204,211] These examples suggest that ML-based basis optimization in nuclear DFT would be useful not only for building adaptive basis representations, but also for choosing basis parameters and estimating basis-truncation effects. To the best of our knowledge, explicitly ML-based construction or optimization of bases has so far not been explored in nuclear DFT-related contexts.

### 3.2.2 Density representation

Density featurization is important to featurize ML functionals. One can distinguish features that encode the entire density distribution or density topology in the vicinity of a point in space when machine learning energy density. When encoding the entire density distribution, a direct grid representation is onerous in electronic DFT. While it has been used in less than three dimensions in both electronic[106,169,174,175] and nuclear[139,142] DFT, it is not feasible or desirable in 3D in the case of electronic DFT (where one wishes to compute systems with large numbers of atoms, and enhancing size capability is a major purpose of ML-enhanced DFT methods). In nuclear DFT, on the contrary, 3D grids are viable for most nuclei. Discrete variable representation has also been used in a 1D electronic problem.[78] Ref. [130] used principal components of promolecular electron density. In Refs. [127,178], coefficients of a harmonic oscillator-like basis functions were used to featurize the electron density. In 3D, one often used a representation of the density in a basis, and basis coefficients then serve as features. Gaussian (with or without spherical harmonics for angular coordinates)[95,103,147,168,212–214] and Fourier[16,177,215] bases have been used for electron density. In Ref. [123] to machine-learn $\tilde{\rho}_p([\rho(\boldsymbol{r})])$ in $(\nabla^2 - \lambda^2)V_p([\rho(\boldsymbol{r})]) = -4\pi\tilde{\rho}_p([\rho(\boldsymbol{r})])$, where $V_p([\rho(\boldsymbol{r})])$ is Pauli potential, features were computed as $d_i = \int_0^{R_c} \rho(\boldsymbol{r}) e^{-r/a_i} sin\left(\frac{\pi m_i r}{R_c}\right) f_c(\boldsymbol{r}) d\boldsymbol{r}$, where $m_i$ and $a_i$ are parameters, $R_c$ is a cutoff radius, and $f_c(\boldsymbol{r})$ is a cutoff function. In Ref. [213], a combination of short-ranged orthogonal Bessel (SR-OB) functions and long-ranged damped-multipole (LR-DM) functions was used for the radial portion of basis functions (with spherical harmonics for the radial part), when machine learning the mapping between the external potential and electron density. with SOAP[161]-based density encoding was used in Refs. [160,162].

Featurization for ML of energy density often involves powers (and sometimes other nonlinear functions[90,124,216]) of the density and its derivatives at points in space, effectively terms of the gradient expansion.[105,114–118,120,122,128,167,184,217] In Refs. [117–120], a case was made for the utility of the product of the electron density and KS effective potential (which can be related to Pauli potential[125]) as a feature. Convolutional filters can be applied to gridded values of the density or density-dependent quantities, further increasing the descriptive power.[121,126]

Both grid and basis representations of the density could be ML-optimized for more efficient featurization for ML functionals. Optimized representations can also be used for more efficient *output* of the density with methods that predict electron density with ML methods from descriptors of chemical composition and structure[218–226] – such predicted densities are important in the context of ML-enhanced DFT in that, when an ML-predicted density is accurate enough, it can be used as is as input to an ML functional, obviating the need to optimize the density and afferent issues with stability with respect to functional derivative. Alternatively, such predicted densities can be used to speed up optimization, initializing the density closer to the optimum. Similar uses on predicted density could be explored in nuclear DFT. There is a relation between the optimality (completeness) of the basis and the necessary number of real-space points sampling it: the more complete the basis the fewer sampling points are needed.[197,227] There is therefore potential in studies of ML-optimized density "grids" and ML-optimized basis sets for representing the density (either electron or nucleon density) that has not need explored yet.

Experimentally, *nuclear* charge densities extracted from elastic electron scattering are usually represented by parameterized forms, such as modified harmonic-oscillator, Fermi, or Gaussian distributions,[228] or by more model-independent expansions such as Fourier-Bessel and sum-of-Gaussians representations,[229,230] i.e. types of bases also often used to represent electron density in electronic DFT. The accuracy and robustness of these approaches depend on the assumed form, cutoff radius, or fitting procedure.[228–230] In nuclear DFT calculations, however, density representations in a basis are practically not used, and nuclear densities are typically stored on spatial grids or reconstructed from occupied single-particle orbitals in mean-field or Kohn-Sham calculations. Grid representations are flexible but may be computationally heavy, whereas orbital summation reintroduces the orbital degrees of freedom that orbital-free DFT seeks to avoid. Therefore, compact and systematic density representations are valuable for nuclear DFT, especially for orbital-free formulations in which the density is the fundamental variable. A recent ML-optimized example of density representation is to use principal component analysis to build a compact orthogonal basis for nuclear density representation, achieving accurate reconstructions with only a few components, more efficiently than traditional methods.[231] Basis representations of nucleon densities, including the use of the bases that are already well understood from experimental studies,[228–230] which could be further improved with ML, thus offers a promising avenue for density featurization for ML-enhanced nuclear DFT. ML-optimized grids may also be promising.

### *3.3 Pseudopotentials, convergence, and other issues*

For electronic KS DFT, there exist sufficiently accurate non-local pseudopotentials of various types,[29–32] and the field of ML PPs is therefore muted. Non-ML optimization of PP parameters is typically performed.[232,233] Nevertheless, methods from the ML toolkit have been used. For example, genetic algorithms (GA) have been used to optimize PAW PPs: in Refs. [234] to optimize the cutoff radius and projector energies for the *s* and *p* orbitals, and in Ref. [235] to optimize the PAW atomic dataset. Such optimizations are most useful for cases where standard PP libraries might not work well, such as the modeling of materials at ultra-high pressure.[235] ML PPs are more in demand for empirical pseudopotential (EP) method[236,237] in which the effective potential of a crystal if fitted to experimental data under a frozen core approximation. As the potential in the EP method is a local potential experienced by one electron, the developments in the ML EP method are relevant for LPP development for OF DFT. The field of ML PPs for the EP methos was initiated in Ref. [238], in which an NN was used to machine-learn a LPP for Si and O in Si and $SiO_2$ solids. The authors used, in particular, SOAP[161] and related features.

ML LPP for OF DFT can be said to be more critically important as availability of good LPPs is a bottleneck preventing the use of electronic OF DFT in applications. LPP tuning often involves fitting parameters of a pre-defined function.[128] ML LPP for electronic OF DFT were also reviewed in Ref. [83]. This field still has a relatively limited body of work but has significant potential. GA was used in Ref. [239] to fit the parameters of analytic expressions of LPPs of Li, Na, and Mg to experimental and KS DFT values of geometries and energies of different phases and mechanical properties. That work also showed that pseudopotential can also be use to effectively palliate KEF deficiencies, in that case to reproduce the cohesive energy whose calculation involves energies of both bulk and atoms, which are not simultaneously accurate with existing functionals (WT[170] and WGC[240]). Beyond finding parameters of an analytic function, proper ML-based representations of LPP have also been considered: a GPR-based representation for Sn pseudopotential was developed in Ref. [241] to reproduce the experimental lattice constants and energy differences between α- and β-Sn phases as well as KS DFT electron density distributions. Rather than using an ML representation of LPP functional form, Ref. [242] used NN to represent norm-conserving pseudo-wavefunctions $\psi_l(r) = N_{nc} r^{l+1} e^{NN_l(r)}$ (corresponding to energies $\epsilon_l$) and the LPP was computed from $\psi_l(r)$ and $\epsilon_l$. This approach resulted in energies, structures, energy-volume curves, and band structures of multiple compounds made of $s$- and $p$-block elements of the second to the fourth periods approaching those computed with non-local pseudopotentials.

Softening of interaction is also helpful for the convergence of self-consistency cycles. The convergence of self-consistency cycles can also be accelerated with ML. Ref. [214] used an equivariant graph transformer NN to emulate SCF cycles, predicting SCF electron density from an input density and descriptors of composition and structure. "Traditional" DFT-like accuracy was achievable on QM9[243] and QMugs[244] molecular datasets.

In nuclear DFT, there is no direct analogue of the core-valence pseudopotential issue encountered in electronic DFT, and correspondingly no existing ML studies devoted to similar topics. However, strategies there for electronic DFT to accelerate convergence of self-consistency cycles could also be useful in nuclear DFT, where a large fraction of the computational cost of large-scale EDF applications comes from repeatedly solving self-consistent mean-field or HFB equations. ML models could be used to predict good initial densities, or to emulate part of the SCF update and mixing procedure. Used in this way, ML would not replace the final variational EDF calculation, but could reduce the number of iterations and lower the cost of large-scale nuclear EDF calculations.

# 4 Conclusions and perspectives: different promising methodological and algorithmic choices for different flavors of DFT

In this perspective, we summarized how ML is used to improve DFT. We focused on four flavors of DFT: electronic and nuclear, Kohn-Sham and orbital-free, tried to make transpire both similarities and differences in how ML/AI methods are or can be used in them, and provided our perspective on promising directions. There are significant opportunities for cross-field fertilization between the fields of electronic and nuclear ML-enhanced DFT. As the field progresses towards more powerful and applications-ready ML-enhanced DFT and from the use of common ML methods towards methods that can be called AI such as formula discovery, we believe such cross-fertilization will be all the more important.

The key ingredient of DFT that is specific to its core is the energy density functional. In electronic DFT, fields of ML XC functionals and KEF are already thriving. Similar ML methods tend to be used for both, and NNs remain the most actively pursued type of method. However, the accuracy and CPU cost requirements are not the same. ML XC functionals are most often developed for KS DFT; their role is to reproduce only the exchange-correlation portion of the total energy – an important yet relatively small part of total energy, on the order of 5-10%. The critical correlation component (as exact exchange can be used to help reproduce exchange energy) is typically as small as 1%. In contrast, the kinetic energy that needs to be reproduced by an ML KEF is as high as 60-70% of total energy. The accuracy requirements are much more stringent then. At the same time, in the KS paradigm, calls to the XC functional are typically not the computational bottleneck, solving a large eigenvalue problem is what is responsible for the near-cubic scaling. This in principle permits using relatively costly ML solutions such as large NNs without introducing large overhead. In OF DFT, on the other hand, the CPU cost of calls to the EDF is a major component of the CPU cost of the calculation, and the use of large NNs could end up partially defeating the purpose. We therefore view lighter approaches including smaller NNs (that, as was indicated above, can have performance on par with move involved algorithms such as GNN), kernel regressions or even low-order polynomial regressions on appropriately defined features, and especially ML-derived formulas as particularly promising for KEFs. ML XC functionals can also be used in OF DFT, where the cost concerns of more involved ML architectures also acquire importance, in which case lighter ML solutions would be desirable. This issue interplays with accuracy: a lighter ML solution can be facilitated by using physically-motivated features or restrictions, in which case the model is expected to embed more error cancellation (important for the calculation of interaction energies) than a large NN.

In nuclear DFT, there is emerging field of ML KEFs that deals with many of the same issues and the field of ML KEFs for electronic DFT. However, the relevant accuracy requirements and practical targets in the description of nuclear ground states differ from those in electronic DFT; moreover, sub-dimensional, e.g., symmetry-constrained calculations, are relevant for applied calculations on real nuclei, while in electronic DFT, sub-dimensional models are mainly useful to test ideas and often use solutions, such as certain grid-based representations, that are not practical in 3D calculations on real materials. This creates a potential for ML-enhanced OF DFT to fast-track in nuclear applications, while electronic ML

OF DFT still faces severe challenges in consistently achieving chemical accuracy beyond simple metals, not only due to the lack of accurate KEFs but also due to the lack of accurate LPPs. Nuclear ML-enhanced OF DFT can and has been benefiting from the experiences gained in electronic DFT. Certain promising approaches worked out in the field of ML KEFs for electronic DFT can benefit nuclear OF DFT in the future, such as the use of the terms of the gradient expansion as input features. There is no direct equivalent to ML XC in nuclear DFT because the underlying nuclear interactions are largely unknown and should be constructed phenomenologically to begin with, as opposed to the known Coulombic interactions in electronic DFT. Nevertheless, the advent of AI methods creates the possibility of discovering key terms in nuclear EDFs in the future.

ML-optimized basis representations of KS states and of the density, for the purpose of achieving compact bases, could be beneficial for both electronic and nuclear DFT. There is a limited number of works to date but good potential for future research. Here also, lighter ML solutions (than complex NN) are desired. While more optimal bases are clearly desired even for medium-scale electronic DFT calculations and are critically important to enable large-scale electronic structure calculations, basis size is often not considered as a critical limiting issue in most nuclear DFT calculations. Nevertheless, ML-optimized bases could be beneficial for highly-accurate nuclear DFT calculations or large-scale calculations over the nuclear chart or for faster sampling of nucleus PESs. As similar types of basis functions found use in electronic and nuclear DFT, cross-fertilization can be fruitful.

Symbolic regression or automatic formular construction has so far been used in a limited way but appears to be promising for electronic and nuclear, KS and OF DFT, and for functionals, basis functions, and pseudopotentials. It marries the expressive power of ML methods with the speed of calculations and portability provided by simple formulas. Importantly, it offers interpretability that can help guide theory development. While symbolic regression is promising for all aspects of ML-enhanced DFT, it is particularly promising in areas where the high CPU cost of ML models can form bottlenecks. This includes KEFs for OF DFT as most of the CPU cost of an OF DFT calculation is due to functional calls. It also includes ML-optimized basis functions, where calling a large ML model for each basis function would be prone to defeat the purpose despite a smaller size of the resulting basis set. For the same reason, direct use of complex ML models for representing KS states[17] does not appear to us particularly promising, as a large number of such states need to be computed.

The compute-heavy nature of complex NN architectures such as convolutional and graph NN that made their way into ML-enhanced DFT can be palliated with GPU- and TPU-amenable algorithms and the use of corresponding hardware. This addresses the underlying issue only partially as GPU and TPU come with their own cost, energy consumption, and availability concerns. NPU (neural processing units)[245] specifically dedicated to accelerating digital neurocomputing could be promising. A quantum leap is promised by neuromorphic computing[246] whereby an NN implementing, for example, a functional and that may be trained on a digital computer is printed on an analog circuit, and ultimately optical computing.[247] However, these technologies are not mature. It also remains open if they are suited for highly accurate ML that is necessary for ML-enhanced DFT, see Refs. [248,249] for the effect of limitations specific to neuromorphic circuits, in particular, in high-accuracy regimes such as ML KEF.[250] The use of simpler and faster, "lighter" ML methods than large NNs still therefore appears to be desired for ML-enhanced DFT and in particular OF-DFT.

The issues of computational cost as well as stability can also be simultaneously palliated by evaluating the EDF on precomputed densities without further optimization. Such densities can be predicted from compositional and structural descriptors via machine learning and other related approaches, as discussed in Refs,. [218–225]. Research targeting electron density prediction is advancing at a fast pace. It is reasonable to expect that sufficiently precise electron densities can be readily obtained for non-self-consistent EDF calculations, at least for a range of material systems. These strategies are also promising for nuclear DFT applications.

## 5 Author contributions

Xin-Hui Wu: Conceptualization, Methodology, Project administration, Writing – original draft, Writing – review & editing.

Sergei Manzhos: Conceptualization, Methodology, Project administration, Writing – original draft, Writing – review & editing.

## 6 Conflicts of interest

There are no conflicts to declare.

## 7 Data availability

Data sharing is not applicable to this article as no new data were created or analyzed in this review.

## 8 Acknowledgements

XHW was partly supported by the National Natural Science Foundation of China under Grant No. 12405134.

We thank Tengxiang Li for providing the image for Figure 5.